\documentclass[aps,prd,reprint]{revtex4-1}

\usepackage{url}
\usepackage{natbib}
\usepackage{textcase}
\usepackage[colorlinks=true,
            linkcolor=blue,
            citecolor=blue]{hyperref}

\usepackage[ngerman, english]{babel}
\usepackage[utf8]{inputenc}
\usepackage{amsmath, amssymb}
\usepackage{slashed}		
\usepackage{braket} 

\usepackage{tabularx} 
\usepackage{colortbl} 
\usepackage{subfigure}
\usepackage{bbold} 

\usepackage{placeins} 
\usepackage{paralist} 

  \usepackage{graphicx}
  \usepackage{epstopdf}	

\def\nn{\nonumber}

\begin{document}
\title{Threshold Resummation for Polarized High-$p_T$ Hadron Production at COMPASS}
\author{Claudia Uebler}
\author{Andreas Schäfer}
\affiliation{Institute for Theoretical Physics, University of Regensburg, D-93053 Regensburg, Germany}
\author{Werner Vogelsang}
\affiliation{Institute for Theoretical Physics, Tübingen University, D-72076 Tübingen, Germany}
\begin{abstract}
We study the cross section for the photoproduction process $\gamma N\rightarrow h X$ 
where the incident photon and nucleon are longitudinally polarized and a hadron $h$ is observed at high transverse 
momentum. Specifically, we address the ``direct'' part of the cross section, for which the photon interacts
in a pointlike way. For this contribution we perform an all-order resummation of logarithmic 
threshold corrections generated by soft or collinear gluon emission to next-to-leading logarithmic accuracy. We present
phenomenological results relevant for the COMPASS experiment and compare to recent COMPASS data.
\end{abstract}

\maketitle
\section{Introduction}
To obtain information about the nucleon's gluon heli\-city distribution $\Delta g$ and to explore its
contribution to the proton's spin is the main focus of several current experiments. One of the probes
employed for this purpose at CERN's COMPASS experiment is $\mu N \rightarrow \mu' h X$, where $h$
denotes a charged hadron produced at high transverse momentum. Kinematics for the 
process are chosen in such a way that the photons exchanged between the muon and the nucleon are
almost real, so that the process effectively becomes $\gamma N \rightarrow h X$. Its double-longitudinal 
spin asymmetry $A_{LL}$ is directly sensitive to $\Delta g$, thanks to the presence of the photon-gluon fusion 
subprocess $\gamma g\to q\bar{q}$. COMPASS has recently presented data for the spin-averaged cross 
section~\cite{Adolph:2012nm} for the process, as well as for its spin asymmetry~\cite{Adolph:2015hta,Levillain:2015twa}.

Thanks to the produced hadron's large transverse momentum, the process $\gamma N \rightarrow h X$
may be treated with perturbative methods. As is well known~\cite{Klasen:2002xb}, hard 
photoproduction cross sections receive contributions from two sources, the ``direct'' ones, for which the photon 
interacts in the usual pointlike way in the hard scattering, and the ``resolved'' ones, for which the photon reveals its own 
partonic structure. Both contributions are of the same order in perturbation theory, starting at ${\cal O}
(\alpha\alpha_s)$, with the electromagnetic and strong coupling constants $\alpha$ and $\alpha_s$. 
Next-to-leading order (NLO, ${\cal O}(\alpha\alpha_s^2)$) QCD corrections for the spin asymmetry for
$\gamma N \rightarrow h X$ have been derived in Refs.~\cite{deFlorian1998} and~\cite{Jaeger2003,Jaeger2005} 
for the direct and resolved cases, respectively. 

As discussed in~\cite{MelaniesPaper}, in the kinematic regime accessible at COMPASS perturbative corrections
beyond NLO are important. This is because typical transverse momenta $p_T$ of the produced hadron are such that the 
variable $x_T = 2p_T/\sqrt{S}$ (with $\sqrt{S}$ the muon-proton center-of-mass energy) is relatively large, $x_T\gtrsim 0.2$. 
This means that the partonic hard-scattering cross sections relevant for $\gamma N \rightarrow h X$ are largely probed in the 
``threshold''-regime, where the initial photon and parton have just enough energy to produce a pair of recoiling high-$p_T$
partons, one of which subsequently fragments into the observed hadron. The phase space for radiation of additional gluons
then becomes small, allowing radiation of only soft and/or collinear gluons. As a result, the cancelation of infrared singularities 
between real and virtual diagrams leaves behind large double- and single-logarithmic corrections to the partonic cross sections. 
These logarithms appear for the first time at NLO and then recur with increasing power at every order of perturbation theory. 
Threshold resummation~\cite{Sterman1987,Catani:1989ne} allows to sum the logarithms to all orders to a certain logarithmic accuracy. 
It was applied to the spin-averaged cross section at COMPASS at next-to-leading logarithm (NLL) level in Ref.~\cite{MelaniesPaper},
where the resummation of both the direct and the resolved contribution was performed. The resummed result for the cross section 
was found to be significantly higher than the NLO one, by roughly a factor two. Comparison to the COMPASS data reported 
in~\cite{Adolph:2012nm} showed that this enhancement is crucial for achieving good agreement between data and the 
perturbative-QCD prediction. 

In the light of this result, it is clear that threshold resummation should also be taken into account in the
theoretical analysis of the spin asymmetry $A_{LL}$ measured at COMPASS~\cite{Adolph:2015hta}.
$A_{LL}$ is the ratio of the spin-dependent cross section and the spin-averaged one. Since the latter has 
already been addressed in~\cite{MelaniesPaper}, we will in this paper examine threshold resummation for 
polarized scattering. As a first step, we will consider the direct contributions to the cross section, which are
simpler to analyze and also formally dominate over the resolved ones near partonic threshold.
We plan to complete our resummation study for $A_{LL}$ in a future publication by
performing threshold resummation also for the resolved contribution in the spin-dependent case. 
We note that the resolved contribution to $\gamma N \rightarrow h X$ is structurally equivalent to
hadronic scattering $pp\to hX$, for which threshold resummation was performed in the previous 
literature even for the polarized case~\cite{deFlorian2007}. However, Ref.~\cite{deFlorian2007} only 
addressed the simplified case when the cross section is integrated over all rapidities of the produced 
hadron, while in the present case we consider an arbitrary fixed rapidity. The techniques necessary
for this were devloped in~\cite{Almeida2009,MelaniesPaper} and will be used here as well. 

Our paper is organized as follows: In Section~\ref{sec2} we recall the general framework for the 
process $\gamma N \rightarrow h X$ in QCD perturbation theory. Section~\ref{sec3} collects all 
ingredients for the threshold resummed spin-dependent cross section. 
In Section~\ref{sec4} we present phenomenological results
for the spin-dependent and spin-averaged cross sections at COMPASS, as well as for the 
resulting longitudinal double-spin asymmetry. Finally, we conclude our paper in Sec.~\ref{sec5}.

\section{Photoproduction cross section in perturbation theory \label{sec2}}
\begin{figure}[t]
      \centering
\vspace*{-1.3cm}
      \includegraphics[width=0.75\linewidth]{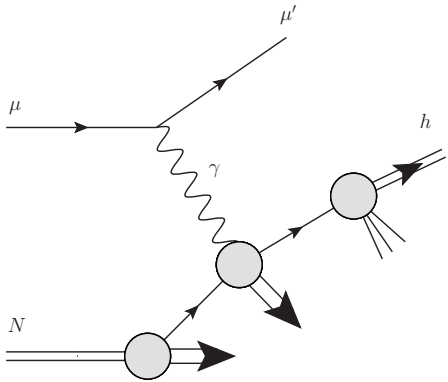}
\vspace*{-1.5cm}
      \caption{\label{lepton-nucleon-dir} {\it High-$p_T$ hadron production in muon-nucleon scattering 
      via direct photon-parton interaction.}}
    \end{figure}

We consider the process
\begin{align}
\ell N \rightarrow \ell' h X\,,
\end{align}
where the lepton $\ell$ and the nucleon $N$ are both longitudinally polarized and where 
a charged hadron $h$ is observed at high transverse momentum $p_T$ (see Fig. \ref{lepton-nucleon-dir}).
Demanding the scattered lepton $\ell'$ to have a low scattering angle with respect to the incoming
one, the main contributions come from almost on-shell photons exchanged between the
lepton and the nucleon. The scattering may then be treated as a {\it photoproduction} process
$\gamma N \rightarrow h X$, with the incoming lepton essentially serving as a source
of quasi-real photons. 

We introduce the spin-averaged and spin-dependent cross sections for the lepton-nucleon process as 
\begin{align}\label{lNhel}
d\sigma_{\ell N} & \equiv \frac{1}{2} \left[ d\sigma_{\ell N}^{++} + d\sigma_{\ell N}^{+-} \right]\,, \nn\\[2mm]
d\Delta\sigma_{\ell N} & \equiv \frac{1}{2} \left[ d\sigma_{\ell N}^{++} -d\sigma_{\ell N}^{+-} \right]\,,
\end{align}
where the superscripts $(++)$, $(+-)$ denote the helicities of the incoming particles. 
Using factorization, the differential spin-dependent cross section (as function of the hadron's
transverse momentum $p_T$ and pseudorapidity $\eta$) may be written 
as \cite{Jaeger2005,MelaniesPaper}: 
\begin{align}
	& \!\!\!\! \frac{p_T^3 d \Delta \sigma}{dp_T d\eta}= \sum_{abc} \int_{x_\ell^{{\mathrm{min}}}}^{1}  \!\!\!\! dx_\ell 
	 \int_{x_n^{{\mathrm{min}}}}^{1} \!\!\!\! dx_n \int_{x}^{1} \!\! dz   \nonumber \\[2mm]
	& \times\frac{\hat{x}_T^4 z^2 }{8v} 
	 \frac{\hat{s} d\Delta\hat{\sigma}_{ab\rightarrow cX}(v,w,\hat{s},\mu_r,\mu_{fi},\mu_{ff})}{dv \, dw}  \nonumber \\[2mm]
	 &\times \Delta f_{a/\ell} \! \left( x_\ell, \mu_{fi} \right) \Delta f_{b/N} \! \left( x_n, \mu_{fi} \right)  D_{h/c} \! \left( z, \mu_{ff} \right) \,,
	 \label{ordinary_factorization}
\end{align}
the sum running over all possible partonic channels $a b \rightarrow c X$. 
The $\Delta f_{b/N} \left( x_n, \mu_{fi} \right)$ are the polarized parton distribution functions of the nucleon, 
which depend on the momentum fraction $x_n$ carried by parton $b$ and on an initial-state factorization scale $\mu_{fi}$. 
They can be written as differences of distributions for positive or negative helicity in a parent nucleon of positive helicity,
\begin{align}
 \Delta f_{b/N}(x,\mu)  \equiv f_{b/N}^+(x,\mu)-f_{b/N}^-(x,\mu)\,.
  \label{spin-depend-gluon-distribution}
\end{align}
In Eq.~(\ref{ordinary_factorization}) we have introduced also ``effective'' parton distributions in a lepton,
 $\Delta f_{a/\ell}  \left( x_\ell, \mu_{fi} \right)$, which we shall elaborate on further below. For now we 
 just note that in terms of spin-dependence they are defined exactly as in~(\ref{spin-depend-gluon-distribution}).
The $D_{h/c}(z, \mu_{ff} )$ in~(\ref{ordinary_factorization}) are the parton-to-hadron fragmentation functions
that describe the hadronization of parton $c$ into hadron $h$, with $z$ being the fraction of the parton $c$'s momentum 
taken by the hadron and $\mu_{ff}$ a final-state factorization scale. Finally, the $d\Delta\hat{\sigma}_{ab\rightarrow cX}$ 
are the spin-dependent cross sections for the partonic hard-scattering processes $ab\rightarrow cX$. In analogy 
with~(\ref{lNhel}) they are defined as
\begin{align}\label{lNhel1}
d\Delta\hat{\sigma}_{ab\rightarrow cX} & \equiv \frac{1}{2} \left[ d\hat{\sigma}_{ab\rightarrow cX}^{++} -
d\hat{\sigma}_{ab\rightarrow cX}^{+-} \right]\,,
\end{align}
the indices now denoting the helicities of the incoming partons. The $d\Delta\hat{\sigma}_{ab\rightarrow cX}$ 
are perturbative and may hence be expanded in terms of the strong coupling constant $\alpha_s$, 
\begin{align}
d\Delta\hat{\sigma}_{ab\rightarrow cX} = d\Delta\hat{\sigma}_{ab\rightarrow cX}^{(0)} + 
\frac{\alpha_s}{\pi}d\Delta\hat{\sigma}_{ab\rightarrow cX}^{(1)} + ... \, .
\end{align}
Apart from the partonic kinematic variables that will be introduced shortly, they depend on the 
factorization scales and also on a renormalization scale $\mu_r$. We note that all formulas presented so far
may be easily written for the spin averaged case by simply summing over helicities 
in~(\ref{spin-depend-gluon-distribution}),(\ref{lNhel1}) instead of taking differences. This then gives the 
unpolarized cross section introduced in Eq.~(\ref{lNhel}) in terms of the usual spin-averaged parton distributions 
$f_{a/\ell},f_{b/N}$ and partonic cross sections $d\hat{\sigma}_{ab\rightarrow cX}$. 

In Eq.~(\ref{ordinary_factorization}) we have introduced a number of kinematic variables. The partonic cross sections 
have been written differential in
\begin{align}\label{vwdef}
 v\equiv 1 +\frac{\hat{t}}{\hat{s}} \quad \text{and} \quad w \equiv \frac{- \hat{u}}{\hat{s}+ \hat{t}}\,,
\end{align}
with the Mandelstam variables 
\begin{align}
 \hat{s}=&(p_a + p_b)^2 = x_\ell x_n S  , \quad \quad \hat{t}=(p_a - p_c)^2= - \frac{\hat{s} \hat{x}_T }{2} e^{- \hat{\eta}}, \nonumber \\
 \hat{u}=&(p_b - p_c)^2= -\frac{\hat{s} \hat{x}_T }{2} e^{\hat{\eta}} ,
\end{align}
where $p_a,p_b,p_c$ are the four-momenta of the participating partons and where 
$S=(p_\ell + p_n)^2$, with the lepton (nucleon) momentum $p_\ell$ ($p_n$). Furthermore, 
\begin{align}
\hat{x}_T\equiv \frac{x_T}{z \sqrt{x_\ell x_n}}\,,
\end{align}
where $x_T\equiv 2 p_T/\sqrt{S}$, and the relationship between the hadron and the parton level center-of-mass 
system rapidities is 
\begin{align}
\hat{\eta}= \eta + \frac{1}{2} \ln \frac{x_n}{x_\ell}\,.
\end{align}
Finally, the lower integration bounds in Eq. (\ref{ordinary_factorization}) are given by
\begin{align}
 x_\ell^{{\mathrm{min}}}&= \frac{x_T e^{\eta} }{2 - x_T e^{- \eta}}, \quad x_n^{{\mathrm{min}}}
 = \frac{x_\ell x_T e^{-\eta} }{2 x_\ell - x_T e^{\eta}}, \nonumber \\[2mm]
x&=\frac{x_T \cosh \hat{\eta}}{\sqrt{x_n x_\ell}}\,.
\end{align}

\begin{figure}[t]
      \centering
      \vspace*{-1.3cm}
      \includegraphics[width=0.75\linewidth]{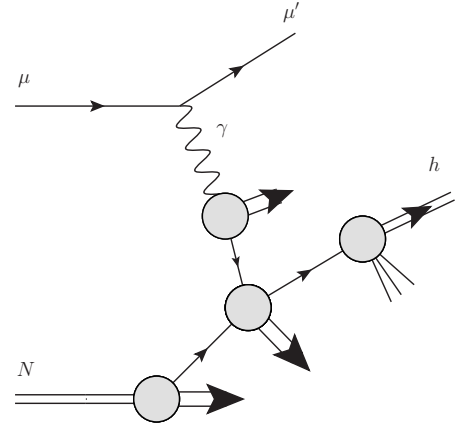}
\vspace*{-1.3cm}
      \caption{\label{lepton-nucleon-resolved} {\it High-$p_T$ hadron production in muon-nucleon scattering via a resolved photon.}}
    \end{figure}

An important aspect of photoproduction cross sections is that the quasi-real photon can interact in two 
ways. For the \textit{direct} contributions (see Fig.~\ref{lepton-nucleon-dir}), 
it participates directly in the hard-scattering, coupling in the
usual pointlike way to quarks and antiquarks. However, as is well established, the photon may also itself
behave like a hadron, revealing its own partonic structure in terms of quarks, antiquarks, and gluons,
as shown in Fig. \ref{lepton-nucleon-resolved}. 
The associated contributions are known as \textit{resolved} photon contributions. The physical 
cross section is the sum of the direct and the resolved part: 
\begin{align}
  d \Delta \sigma =  d \Delta \sigma_{{\mathrm{dir}}} +  d \Delta \sigma_{{\mathrm{res}}}\,.
\end{align}
Both contributions are captured by Eq.~(\ref{ordinary_factorization}) by introducing an
effective spin-dependent parton distribution for the lepton:
\begin{align}
  \Delta f_{a/\ell} (x_\ell, \mu_{ff}) = \int_{x_\ell}^{1} \frac{dy}{y} 
  \Delta P_{\gamma \ell}(y) \Delta f_{a/ \gamma} \left(x_\gamma = \frac{x_\ell}{y}, \mu_{ff}\right) \,,
  \label{PDF_lepton}
\end{align}
where $\Delta P_{\gamma \ell } (y)$ is the polarized Weizs\"{a}cker-Williams spectrum and
$\Delta f_{a/ \gamma}$ describes the distribution of parton $a$ inside the photon. 
Equation~(\ref{PDF_lepton}) also applies to the direct case, see Fig. \ref{lepton-nucleon-dir}, 
where parton $a$ is an elementary photon and hence
\begin{align}
 \Delta f_{\gamma / \gamma}= \delta \left( 1 - x_{\gamma}\right)\,.
\end{align}
The Weizs\"{a}cker-Williams spectrum is given by~\cite{deFlorian:1999ge}:
\begin{align}
 \Delta P_{\gamma \ell} (y) =& \frac{\alpha}{2\pi} 
 \left[ \frac{1 - \left(1 - y \right)^2 }{y} \ln{ \left( \frac{ Q_{{\mathrm{max}}}^2 (1-y)}{m_\ell^2 y^2} \right)} \right. 
 \nonumber \\[2mm]
  &  \hspace{6mm} \left.  + 2 m_\ell^2  y^2 \left( \frac{1}{Q_{{\mathrm{max}}}^2} - \frac{1-y}{m_\ell^2 y^2} \right) \right] \,,
 \label{WeizsaeckerSpectrum}
\end{align}
where $\alpha$ is the fine structure constant. $\Delta P_{\gamma \ell}$ describes the (nearly) collinear emission 
of a polarized photon with momentum fraction $y$ by a polarized lepton with mass 
$m_\ell$. The low virtuality $Q^2$ of the photon is restricted by an upper limit $Q_{{\mathrm{max}}}^2$ 
that is determined by the experimental conditions. 

In the direct case, there are three different LO subprocesses, 
\begin{align}
\gamma q \rightarrow g (q) , \quad \gamma q \rightarrow q (g) , \quad \gamma g \rightarrow q (\bar{q}) \,,
 \label{direct_subprocesses}
\end{align}
where the final-state particle in brackets is understood to remain unobserved, while the other parton 
fragments into the observed hadron. We note that the photon-gluon-fusion process is symmetric under 
exchange of $q$ and $\bar{q}$ in the final state. The spin-dependent cross sections for the LO 
subprocesses are \cite{deFlorian1998}:
\begin{align}
 \frac{\hat{s} d \Delta \hat{\sigma}_{\gamma q \rightarrow g (q)}^{(0)} (v,w) }{dvdw} &= 2 \pi \alpha \alpha_s  e_q^2
 C_F \frac{1-v^2}{v} \delta \left( 1-w \right), \displaybreak[1]  \nonumber \\
 \frac{\hat{s} d \Delta \hat{\sigma}_{\gamma q \rightarrow q (g)}^{(0)} (v,w) }{dvdw} &= 2 \pi \alpha \alpha_se_q^2 
 C_F \frac{1-(1-v)^2}{1-v} \delta \left( 1-w \right), \displaybreak[1] \nonumber \\
 \frac{\hat{s} d \Delta \hat{\sigma}_{\gamma g \rightarrow q (\bar{q})}^{(0)} (v,w) }{dvdw} &= - 2 \pi \alpha \alpha_s  e_q^2
T_R  \frac{v^2 + (1-v)^2}{v(1-v)} \delta \left( 1-w \right),
 \label{pol_Born}
\end{align}
with $C_F=4/3$, $T_R=1/2$ and the fractional electromagnetic charge $e_q$ of the quark.

In the resolved case, all $2\to 2$ QCD partonic processes contribute at LO: 
\begin{align}\label{qcd}
  &q\,q' \rightarrow q \, q', \quad  q \, \bar{q}' \rightarrow q  \,\bar{q}', \quad  q \, \bar{q}  
  \rightarrow q' \, \bar{q}', \quad   q \,q \rightarrow q \,q , \nonumber \\
  & q \, \bar{q} \rightarrow q \bar{q} , \quad q \, \bar{q} \rightarrow g \,g, \quad  g \, 
  q \rightarrow q  \,g, \quad  g \, g \rightarrow g \,q, \nonumber \\
  & g \,g \rightarrow g \,g , \quad  g \,g \rightarrow q  \,\bar{q} ,
\end{align}
where either of the final-state partons may fragment into the observed hadron. As the 
photon's parton distributions $\Delta f_{a/ \gamma}$ are of order $\alpha/ \alpha_s$, both the 
direct and the resolved LO contributions are of order $\alpha^2 \alpha_s$ for the 
$\ell N$ cross section (or of order $\alpha \alpha_s$ for the $\gamma N$ one)~\cite{Klasen:2002xb}.
Note that the $\Delta f_{a/ \gamma}$ contain a perturbative ``pointlike'' contribution that dominates
at high $x_\gamma$, but also a nonperturbative ``hadronic'' piece that is associated with the
photon converting into a vector meson and is important at low-to-mid $x_{\gamma}$. 

As shown in Eq.~(\ref{pol_Born}), the LO partonic cross sections are proportional to 
$\delta(1-w)$. From~(\ref{vwdef}) one finds that the invariant mass squared of the final state 
that recoils against the fragmenting parton is given by 
\begin{align}
 \hat{s}_4= \hat{s}+ \hat{t} +\hat{u}=\hat{s} v (1-w) = \hat{s} (1-\hat{x}_T \cosh \hat{\eta}).
 \label{v_w_s4}
\end{align}
The $\delta(1-w)$ at LO thus reflects the fact that the recoil consists of a single massless parton.
At NLO, the partonic cross sections contain various types of distributions in $(1-w)$. 
Analytical expressions have been obtained in 
Refs.~\cite{deFlorian1998,Jaeger2003,Jaeger2002,Aversa:1988vb,Gordon:1994wu,Hinderer:2015hra}.
For each process the result may be cast into the form
\begin{align}
 \frac{\hat{s} d \Delta \hat{\sigma}^{(1)}_{a b\rightarrow c X} (v,w) }{dv \, dw}
 =& A(v) \delta(1-w) + B(v) \left( \frac{\ln(1-w)}{1-w} \right)_{+} \nonumber \\
  & + C(v) \left( \frac{1}{1-w} \right)_{+} + F(v,w) \text{,}
 \label{NLO_shaped}
\end{align}
where the coefficients $A(v),B(v),C(v),F(v,w)$ depend on the process under consideration,
and where the plus-distributions are defined as usual by
\begin{align}
 \int^1_0  dw \, f(w) \left[ g(w) \right]_{+} \equiv  \int^1_0  dw \,\left[  f(w) - f(1) \right] g(w) .
\end{align}
The function $F(v,w)$ in~(\ref{NLO_shaped}) contains all remaining terms without distributions in $(1-w)$. 
The terms with plus-distributions give rise to the large double-logarithmic corrections that are addressed
by threshold resummation. Their origin lies in soft-gluon radiation, and they recur with higher power 
at every higher order of perturbation theory. For the $k$-th order QCD correction, the leading terms are 
proportional to $\alpha_s^k [\ln^{2k-1} (1-w) / (1-w)]_{+}$ (not counting the overall power of the partonic
process in $\alpha_s$). Subleading terms are down by one or more powers of $\ln(1-w)$.

Both the direct and the resolved contributions have the structure shown in~(\ref{NLO_shaped}).
In the following we discuss the all-order resummation of the threshold logarithms in the direct part of the
cross section, which we separate from the resolved part adopting the $\overline{{\mathrm{MS}}}$ scheme.
We perform the resummation to next-to-leading logarithm (NLL), which means that the three 
``towers'' $\alpha_s^k  [\ln^{2k-1} (1-w) / (1-w)]_{+}$, $\alpha_s^k  [\ln^{2k-2} (1-w) / (1-w)]_{+}$, $\alpha_s^k  
[\ln^{2k-3} (1-w) / (1-w)]_{+}$ are taken into account to all orders in the strong coupling.

\section{Resummed cross section \label{sec3}}
\subsection{Transformation to Mellin moment space}
The resummation may be organized in Mellin moment space. A particularly convenient way developed 
in~\cite{Almeida2009,MelaniesPaper} is to start from Eq.~(\ref{ordinary_factorization}) and write the convolution 
of the partonic cross sections with the fragmentation functions as the Mellin inverse of the corresponding products of 
Mellin moments. For the direct contributions we have
\begin{align}
 	 \!\! \frac{p_T^3 d\sigma}{dp_T d\eta}= \sum_{bc} & \int_{0}^{1}  \!\! dx_\ell \int_{0}^{1} \!\! dx_n 
	 \Delta f_{\gamma/\ell}  \left( x_\ell, \mu_{fi} \right) \Delta f_{b/N}  \left( x_n, \mu_{fi} \right) \nonumber \\[2mm]
	 & \times \int_{\mathcal{C}} \frac{dN}{2 \pi i} (x^2)^{-N} D_{h/c}^{2N+3} (\mu_{ff}) 
	 \Delta \tilde{w}_{\gamma b\rightarrow cX}^{2N} \left( \hat{ \eta} \right),
	 \label{factorization}
\end{align}
where 
\begin{align}
D_{h/c}^N (\mu) \equiv \int_0^1 dz \, z^{N-1} D_{h/c} (z, \mu)
\end{align}
and
\begin{align}
 \Delta \tilde{w}_{\gamma b\rightarrow cX}^{N} (\hat{\eta}) \equiv 
 2 \! \! \int^1_0 d \frac{\hat{s}_4}{\hat{s}}  \left(\! 1 -\frac{\hat{s}_4}{\hat{s}}  \right)^{ \! N-1}  \! \frac{\hat{x}_T^4 z^2}{8v} 
 \frac{\hat{s}d\Delta\hat{\sigma}_{\gamma b\rightarrow cX}}{dv \, dw} \text{,}
 \label{hard_scattering_Mellin}
\end{align}
with $\hat{s}_4$ as defined in~(\ref{v_w_s4}). For simplicity, we have not written out the dependence of the
$\Delta \tilde{w}_{\gamma b\rightarrow cX}^N$ on $\hat{s}$ and on the factorization and renormalization scales, 
which they inherit from the $d\Delta\hat{\sigma}_{\gamma b\rightarrow cX}$. As one can see, in writing the cross 
section in the form~(\ref{factorization}) we keep the parton distribution functions in $x$-space. 

The plus-distributions in $(1-w)$ in the $d\Delta\hat{\sigma}_{\gamma b\rightarrow cX}$ turn into logarithms of
the Mellin variable $N$ in the $\Delta \tilde{w}_{\gamma b\rightarrow cX}^N$. Specifically, the terms  
$\alpha_s^k  [\ln^{2k-1} (1-w) / (1-w)]_{+}$, $\alpha_s^k  [\ln^{2k-2} (1-w) / (1-w)]_{+}$, $\alpha_s^k  
[\ln^{2k-3} (1-w) / (1-w)]_{+}$ mentioned above turn into the NLL towers $\alpha_s^k \ln^{2k}(N)$,  $\alpha_s^k 
\ln^{2k-1}(N)$,  $\alpha_s^k \ln^{2k-2}(N)$ in moment space. Threshold resummation provides closed
expressions for the $\Delta \tilde{w}_{\gamma b\rightarrow cX}^N$ that contain these logarithms to all orders.
Inserting these expressions into~(\ref{factorization}) and performing the inverse Mellin transformation and
the convolution with the parton distribution functions then yields the resummed hadronic cross section. 
We note that the presence of the moments of the fragmentation functions in~(\ref{factorization}) is
important for making the Mellin-inverse sufficiently well-behaved that the convolution with the 
parton distribution functions can be carried out numerically. The reason is that the $D_{h/c}^N$
fall off rapidly at large $N$ and thus tame the logarithms in $N$ and hence the plus-distributions
in $(1-w)$.

\subsection{NLL-resummed hard-scattering function}

The resummed expressions for the $\Delta \tilde{w}_{\gamma b\rightarrow cX}^N$ may be 
obtained~\cite{MelaniesPaper} from the corresponding ones for the {\it production} of photons, 
$ab\to \gamma X$, which were derived and discussed in detail 
in~\cite{LaenenJuni1998,Sterman2000,Catani1998}. To NLL, one finds:
\begin{align}
    \Delta & \tilde{w}_{\gamma b \rightarrow cd}^{N,\text{resum}} \left(\hat{\eta} \right) = 
    \left( 1 + \frac{\alpha_s}{\pi} \Delta C^{(1)}_{\gamma b \rightarrow cd} \right) 
     \Delta \hat{\sigma}_{\gamma b \rightarrow cd}^{(0)} (N,\hat{\eta})
    \nonumber \\[2mm]
    & \hspace{0.5cm} \times  \Delta_b^{(-\hat{t}/\hat{s})N} (\hat{s},\mu_{fi},\mu_r)
    \Delta_c^{N} (\hat{s},\mu_{ff},\mu_r) J_d^{N}(\hat{s}) 
    \nonumber \\[2mm]
    & \hspace{0.5cm} \times \exp \left[ \int_{\mu_r}^{\sqrt{\hat{s}}/N} \frac{d \mu'}{\mu'} 
    2 \text{Re} \Gamma_{\gamma b \rightarrow cd} (\hat{\eta},\alpha_s (\mu') ) \right] .
\label{direct_w}
\end{align}
We now discuss the various functions appearing in this expression. We first note that 
among them only $\Delta \hat{\sigma}_{\gamma b \rightarrow cd}^{(0)}$ and
$\Delta C^{(1)}_{\gamma b \rightarrow cd}$ depend on the polarizations
of the incoming partons; all other factors are spin-independent. The $\Delta \hat{\sigma}_{\gamma 
b \rightarrow cd}^{(0)}$ are the Mellin-moments of the Born cross sections in~(\ref{pol_Born}):
\begin{align}
\! \Delta \hat{\sigma}_{\gamma b \rightarrow c d}^{(0)} \! \left( N, \hat{\eta} \right) \! \equiv 2 \! \int_{0}^{1} \! 
d  \frac{\hat{s}_4}{\hat{s}}  \left(\! 1 \! - \! \frac{\hat{s}_4}{\hat{s}} \! \right)^{\! \!N-1} 
 \! \frac{\hat{x}_T^4 z^2}{8v} \frac{\hat{s} d \Delta \hat{\sigma}_{\gamma b \rightarrow c d }^{(0)}}{dv dw} .
\end{align}
We can easily compute them in closed form by exploi\-ting the $\delta(1-w)$-function in~(\ref{pol_Born})
and the relation $\hat{s}_4=\hat{s}v (1-w)$, Eq. (\ref{v_w_s4}). The coefficients $\Delta C^{(1)}_{\gamma b \rightarrow cd}$ 
match the resummed cross section to the NLO one. They correspond to hard contributions and 
primarily originate from the virtual corrections at NLO and may be extracted by comparing the exact NLO 
cross section with the first-order expansion of the resummed one. We have followed this procedure;
our results are given in Appendix~\ref{AppA}. We note that the $\Delta C^{(1)}_{\gamma b \rightarrow cd}$ 
are functions of $v$ and the ratios $\mu^2/\hat{s}$, where $\mu$ is any of the scales $\mu_r,\mu_{fi},\mu_{ff}$.

The functions $\Delta_b^{(-\hat{t}/\hat{s})N}$ and $\Delta_c^{N}$ in~(\ref{direct_w})
account for soft radiation collinear to the initial-state parton $b$ or to the fragmenting parton
$c$, respectively. They are exponentials and given in the $\overline{{\mathrm{MS}}}$ scheme 
as~\cite{LaenenJuni1998}
\begin{align}
  &  \ln \Delta_i^{N}\left(\hat{s},\mu_f,\mu_r \right)=  \nonumber \\[2mm]
    &  - \int_0^1 dz \frac{z^{N-1}-1}{1-z} \int_{(1-z)^2}^1 \frac{dt}{t} A_i\left(\alpha_s(t \hat{s})   \right)  
    \nonumber \\[2mm]
    &  -2 \int_{\mu_r}^{\sqrt{\hat{s}}} \frac{d\mu'}{\mu'} \gamma_i (\alpha_s(\mu'^2))
    + 2 \int_{\mu_{fi}}^{\sqrt{\hat{s}}} \frac{d\mu'}{\mu'} \gamma_{ii} (N, \alpha_s(\mu'^2)), 
\label{ln_Delta}
\end{align}
where the functions $A_i,  \gamma_i ,\gamma_{ii}$ ($i=q,g$) are perturbative series
in the strong coupling that are well-known. For convenience, we collect them 
in Appendix~\ref{AppB}. The function $J_d^{N}$ describes
collinear emission, soft and hard, off the unobserved recoiling parton $d$. We have~\cite{LaenenJuni1998}
\begin{align}
&    \ln J_d^{N}(\hat{s},\mu_r) =\int_0^1 dz \frac{z^{N-1}-1}{1-z} \bigg\{ \nn\\[2mm]
&  \int_{(1-z)^2}^{(1-z)}\frac{dt}{t}  A_d \left(\alpha_s(t \hat{s})\right) - \gamma_d \left(\alpha_s((1 \! -z)\hat{s} )\right)  
		\bigg\} \nonumber \\[2mm]
		&\! + 2 \int_{\mu_{r}}^{\sqrt{\hat{s}}} \frac{d\mu'}{\mu'} \gamma_d \left(\alpha_s(\mu'^2)\right).
	\label{ln_J}	
\end{align}

Finally, emission of soft gluons at large angles is accounted for by the last factor in~(\ref{direct_w}). The
soft anomalous dimension $\Gamma_{\gamma b \rightarrow cd}$ in its exponent starts at 
$\mathcal{O}(\alpha_s)$~\cite{LaenenJuni1998},
\begin{align}
  \Gamma_{\gamma b\rightarrow cd} (\hat{\eta}, \alpha_s )  = \frac{\alpha_s}{\pi} \Gamma_{\gamma 
  b\rightarrow cd}^{(1)} (\hat{\eta}) + \mathcal{O}(\alpha_s^2).
\end{align}
As indicated, it explicitly depends on the pseudo rapidity $\hat\eta$. The first-order terms of the 
anomalous dimensions for our various direct subprocesses can be obtained~\cite{MelaniesPaper} 
from those for the prompt-photon production processes, $q \bar{q}\rightarrow \gamma g$ 
and $q g \rightarrow \gamma q$, given in~\cite{LaenenJuni1998}:
\begin{align}
 \Gamma_{\gamma q \rightarrow q g}^{(1)} \left( \hat{\eta} \right) = &
 C_F \ln{ \left( \frac{-\hat{u}}{\hat{s}} \right) } 
  + \frac{C_A}{2} \left[ \ln{ \left( \frac{\hat{t}}{\hat{u}} \right) } - i\pi   \right] \text{,} \\[2mm]
 \Gamma_{\gamma q \rightarrow g q}^{(1)} \left( \hat{\eta} \right) = & 
 \Gamma_{\gamma q \rightarrow q g}^{(1)} \left( \hat{\eta} \right) \left|_{\hat{t}  
 \leftrightarrow \hat{u}} \right. \text{,}\\[2mm]
 \Gamma_{\gamma g \rightarrow q \bar{q}}^{(1)} \left( \hat{\eta} \right) = &
 C_F\,i\pi  
  + \frac{C_A}{2} \left[ \ln{ \left( \frac{\hat{t} \hat{u}}{\hat{s}^2} \right) }  + i\pi  \right] .
\end{align}
We note that the imaginary parts do not contribute since the real part is taken in the last
exponent in~(\ref{direct_w}).

After inserting all factors into Eq.~(\ref{direct_w}), our final resummed expression 
is obtained by expanding to NLL. The techniques for this are standard, and we present
the results of the expansion in Appendix~\ref{AppB}. We have checked that upon further
expansion of the results to NLO, all single- and double-logarithmic terms of the exact NLO 
partonic cross sections given in~\cite{deFlorian1998} are recovered. The terms constant in $N$ also match 
provided we use the coefficients $\Delta C^{(1)}_{\gamma b \rightarrow cd}$ as given in Appendix~\ref{AppA}.

We finally note that for the direct contributions that we consider in this paper, the LO hard-scattering
cross sections only possess a single color structure, given by that of the $q\bar{q}g$ vertex. Due to 
color conservation, soft-gluon emission thus cannot lead to color transitions in the hard-scattering subprocesses.
This changes when one considers the resolved-photon contributions, for which at LO the $2\to 2$ QCD 
scattering processes in~(\ref{qcd}) contribute. As is 
well known~\cite{deFlorian2007,KidonakisJan1998,Bonciani:2003nt,deFlorian2005,Almeida2009,Hinderer:2014qta},
in this case a matrix structure arises in the resummed cross section. We plan to address the resummation of
the resolved-photon contributions in a future publication. We note that they are formally suppressed by $1/N$ 
relative to the direct ones near threshold, due to the photon's parton distributions. As a result, they
fall off more rapidly toward higher transverse momenta, as we shall see below.

\subsection{Inverse Mellin transform and matching procedure}

\begin{figure}[t]
  \centering
\vspace*{-1.1cm}
  \includegraphics[width=0.75\linewidth]{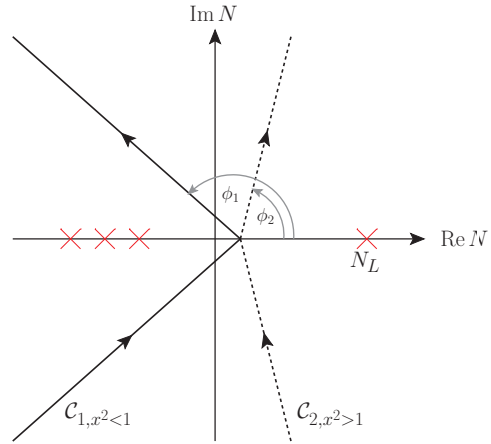}
\vspace*{-1.1cm}
  \caption{\label{Mellin_inverse} {\it Our choices for the contours of the inverse Mellin transform: 
  $\mathcal{C}_1$ for $x^2<1$ and $\mathcal{C}_2$ for $x^2\geq 1$.}}
\end{figure}

As seen in Eq.~(\ref{direct_w}), we need to perform an inverse Mellin transform in order to arrive
at the resummed hadronic cross section. In the course of this we need to deal with singularities appearing 
in the NLL expanded exponents, Eq. (\ref{NLL_exponents1}),(\ref{NLL_exponents2}), 
at $\lambda=1/2$ and $\lambda=1$, where $\lambda=\alpha_s b_0 \ln (N{\mathrm{e}}^{\gamma_E})$.
These singularities are a consequence of the Landau pole in the perturbative strong coupling and 
lie on the positive real axis in moment space. The left of these poles is located at 
$N_L=\exp \left(1/(2\alpha_s b_0)-\gamma_E\right)$ in the complex-$N$ plane. We will use the {\it Minimal Prescription} 
formula introduced in \cite{Catani_MinimalPrescr}, for which one chooses the integration contour
as shown in Fig.~\ref{Mellin_inverse}. The main feature of the contour is that it intersects the real axis
at a value $C_{{\mathrm{MP}}}$ that lies to the {\it left} of $N_L$ (but, of course, to the right of all other
poles originating from the fragmentation functions). 

It is important to point out that the Mellin-integral in~(\ref{direct_w}) defined in this way,
\begin{align}
 \int_{C_{{\mathrm{MP}}}-i\infty}^{C_{{\mathrm{MP}}}+i\infty} \frac{dN}{2 \pi i} (x^2)^{-N} 
 D_{h/c}^{2N+3}(\mu_{ff}) \Delta \tilde{w}^{2N,\text{resum}}(\hat{\eta}),
\end{align}
has support for both $x^2<1$ and $x^2\geq 1$. The latter contributions arise only because
of the way the Landau poles are treated in the Minimal Prescription. They are unphysical in the
sense that the cross section at any {\it finite} order of perturbation theory must not receive any
contributions from $x^2\geq 1$. Mathematically, however, the unphysical contributions
are needed to make sure that the expansions of the resummed cross section to higher orders
in $\alpha_s$ converge to the fully resummed result. We note that the piece with $x^2\geq 1$ 
decreases exponentially with $x^2$, so that its numerical effect is suppressed. As shown in 
Fig.~\ref{Mellin_inverse}, we tilt the contours with respect to the real axis, which helps to 
improve the numerical convergence of the Mellin integral. For $x^2<1$ ($x^2\geq 1$), we need 
to choose an angle $\phi_1>\pi/2$ ($\phi_2<\pi/2$). 

We finally note that as usual we match our resummed cross section to the NLO one by subtracting all
NLO contributions that are present in the resummed result and adding instead the full NLO cross section:
\begin{align}
   & \frac{p_T^3 \Delta d \hat{\sigma}^{\text{matched}} }{dp_T \, d\eta}=  \frac{p_T^3 d \Delta \sigma^{\text{NLO} } }{dp_T \, d\eta}
   +\sum_{bc} \int_0^1 dx_\ell \int_0^1 dx_n 
   \nonumber \\[2mm]
   & \hspace{0.2cm} \times \Delta  f_{\gamma/\ell}(x_\ell,\mu_{fi})\Delta f_{b/N}(x_n,\mu_{fi}) \nonumber \\[2mm]
   & \hspace{0.2cm} \times \int_{\mathcal{C}} \frac{dN}{2 \pi i} (x^2)^{-N} D_{h/c}^{2N+3} (\mu_{ff}) \nonumber \\[2mm]
   & \hspace{0.2cm} \times \left[ \Delta \tilde{w}_{\gamma b \rightarrow cd}^{2N,\text{resum}}(\hat{\eta})  -  \left. 
   \Delta \tilde{w}_{\gamma b \rightarrow cd}^{2N,\text{resum}}(\hat{\eta}) \right|_{\mathrm{NLO}} \,\right] ,   
   \label{final_matched}
\end{align}
where ``$|_{\mathrm{NLO}}$'' denotes the truncation at NLO. This procedure makes sure that NLO is fully included 
in the theoretical predictions, as well as all soft-gluon contributions beyond NLO to NLL accuracy. It avoids any 
double-counting of perturbative terms. 

\section{Phenomenological results \label{sec4}}
As discussed in the Introduction, measurements of cross sections and spin asymmetries for
the photoproduction process $\mu N\rightarrow h X$ are carried out in the COMPASS 
experiment~\cite{Adolph:2012nm,Adolph:2015hta} at CERN. We therefore present our 
phenomenological results for COMPASS kinematics. COMPASS uses a longitudinally polarized 
muon beam with mean beam energy of $E_{\mu}=160$ GeV, resulting in $\sqrt{S}=17.4$ GeV. 
Both deuteron and proton targets are available. COMPASS imposes the cut $Q^2_{{\mathrm{max}}}=
1$ GeV$^2$ on the virtuality of the exchanged photon which we use in the Weizs\"{a}cker-Williams 
spectrum~(\ref{WeizsaeckerSpectrum}). As in COMPASS, we also implement the cuts 
$0.2 \leq y \leq 0.9$ on the fraction of the lepton's momentum carried by the photon, and
$0.2 \leq z \leq 0.8$ for the fraction of the energy of the virtual photon carried by the 
hadron. Finally, charged hadrons are detected in COMPASS if their scattering angle is between 
$10 \leq \theta \leq 120$ mrad, corresponding to $-0.1\leq \eta \leq 2.38$ in the hadron's
pseudorapidity. We integrate over this range. 

Our default choice for the helicity parton distributions is the set of \cite{deFlorian:2014yva} (referred to as DSSV2014). 
We adopt the fragmentation functions of Ref.~\cite{fDSS} (DSS) throughout this work. In the calculations of the NLL 
resummed unpolarized cross sections we follow Ref.~\cite{MelaniesPaper} and use the numerical code of that work. 
Unless stated otherwise, we employ the unpolarized parton distribution functions of Ref.~\cite{Martin:2009iq} 
(referred to as MSTW).  For comparisons we will also present results for the NLO resolved contributions,
for which we will adopt the unpolarized and polarized photonic parton distributions of Refs.~\cite{Gluck:1991jc}
and~\cite{Gluck:1992zq}, respectively. We furthermore choose all factorization/renormalization scales 
to be equal, $ \mu_r = \mu_{fi}= \mu_{ff}\equiv \mu$. We usually choose $\mu=p_T$, except when investigating the 
scale dependence of the theoretical predictions.

\subsection{Polarized and unpolarized resummed cross sections}
\begin{figure}[t]
  \centering
  \includegraphics[width=\linewidth]{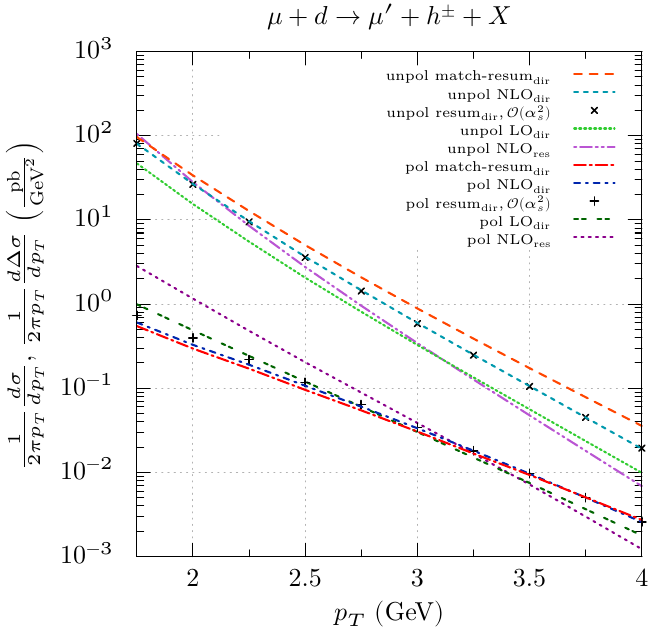}
  \caption{\label{compare_unpol-pol} {\it Direct parts of the spin averaged and spin 
  dependent LO, NLO and matched resummed differential cross sections for 
  $\mu d \rightarrow \mu' h^{\pm} X$. We also show the NLO expansions
  of the resummed results (symbols), as well as the NLO resolved contributions.}}
\end{figure}

Figure~\ref{compare_unpol-pol} shows the direct parts (defined in the $\overline{{\mathrm{MS}}}$ scheme) 
of the spin-averaged and spin-dependent cross sections for $\mu d \rightarrow \mu' h^{\pm} X$ at 
leading order (LO), next-to-leading order, and resummed with matching implemented as described in 
Eq.~(\ref{final_matched}). The symbols in the figure show the NLO-expansions of the non-matched 
resummed cross sections, and for comparison the figure also presents the NLO resolved contributions. 
We have summed over the charges of the produced hadrons. As can be seen, in the unpolarized case 
the difference between the LO and NLO results is very large, and resummation adds another equally 
sizable correction that increases relative to the NLO result as one goes to larger $p_T$, that is,
closer to threshold. The NLO expansion of the resummed cross section shows excellent agreement with the 
full NLO result, demonstrating that the threshold terms correctly reproduce the dominant part of the 
cross section. These findings are as reported in~\cite{MelaniesPaper}. 

In the polarized case, the higher-order corrections are overall much more modest. The NLO prediction
is slightly lower than the LO one at $p_T\lesssim 2.5$~GeV but higher for larger values of transverse
momentum. The resummation effects are smaller here, leading to only a modest further enhancement 
over NLO as one gets closer to threshold. This implies that the higher-order resummation effects
will not cancel in the spin asymmetry for the process. Again the NLO expansion of the resummed cross 
section reproduces the full NLO result faithfully, although not quite as well as in the unpolarized case.
These features that we observe for the direct part of the polarized cross section may be understood
from the fact that the two competing LO subprocesses $\gamma q\to qg$ and 
$\gamma g\to q\bar{q}$ enter with opposite sign and thus cancel to some extent. This 
was already observed in Ref.~\cite{Jaeger2005} in the context of the NLO calculation. 
As discussed there, the cancelation is also responsible for the fact that the resolved contribution 
to the cross section computed with the ``maximal'' set of~\cite{Gluck:1992zq}
is relatively much more important than in the unpolarized case, 
as is evident from the curves for the resolved part shown in the figure. Even though the
resolved contributions also have gluon-initiated subprocesses and hence are sensitive
to $\Delta g$, they have significant uncertainty due to the fact that very little is known about
the spin-dependent parton distributions of the photon. The (possible) dominance of the resolved
contributions in the polarized case thus sets a severe limitation for extractions of $\Delta g$ from
$\gamma N \rightarrow h X$~\cite{Jaeger2005}. 
\begin{figure}[t]
  \centering
  \includegraphics[width=\linewidth]{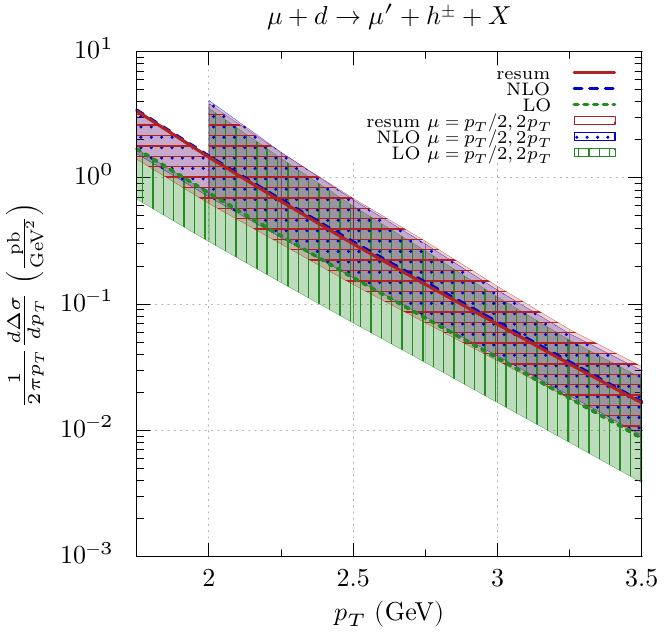}
  \caption{\label{scale-dependence} {\it Scale dependence of the spin-dependent cross section at
  LO, NLO, and for the resummed case. For the resummed cross section we include the resolved contributions at NLO. 
  We vary the scale $\mu= \mu_r = \mu_{fi}= \mu_{ff}$ in the range $p_T/2 \leq \mu \leq 2 p_T $.
  The upper ends of the bands correspond to $\mu= p_T/2$, the lower ones to $\mu= 2 p_T$. 
  We show results only when the scale $\mu$ exceeds 1~GeV.}}
\end{figure}

In Fig.~\ref{scale-dependence} we examine the scale dependence of the spin-dependent cross section.
For the resummed cross section we include the resolved contributions at NLO level, so that
\begin{align}
  d\Delta \sigma_{\text{resum}} = d\Delta \sigma_{\text{dir,resum}} + d\Delta\sigma_{\text{res,NLO}}.
\end{align}
The LO and NLO cross sections contain as usual their full direct and resolved contributions.
We vary the scales in the range $p_T/2 \leq \mu \leq 2 p_T $. One can observe
that the scale uncertainty is large, especially so at the lower $p_T$. There is a clear improvement when going from LO
to NLO, but no further improvement when we include resummation. If anything, the resummed result shows a slightly
larger scale dependence than the NLO one, a feature that will require further attention in the future. 

\subsection{Double-Spin Asymmetry}
\begin{figure}[t]
  \begin{center}
  \subfigure[]{\includegraphics[width=\linewidth]{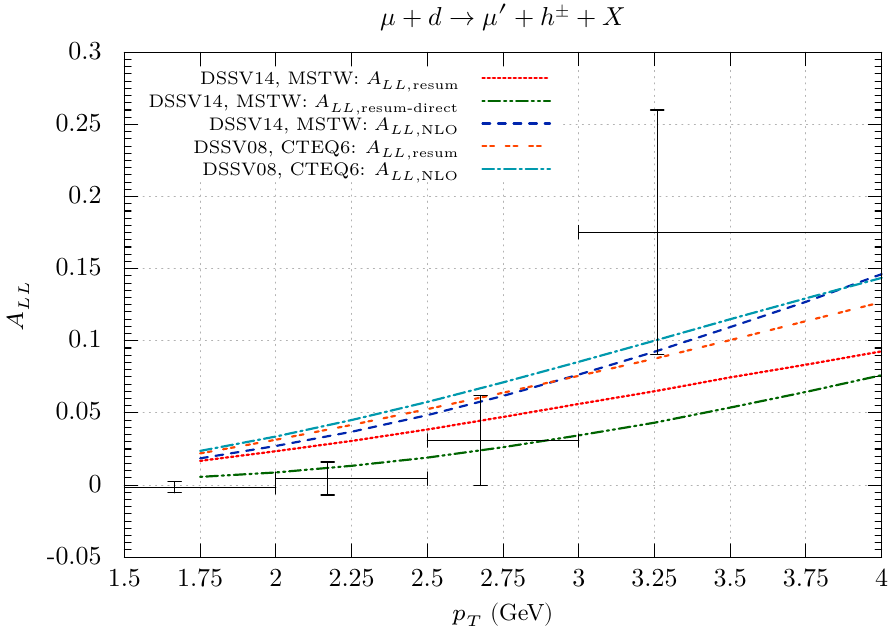}}\\ 
  \subfigure[]{\includegraphics[width=\linewidth]{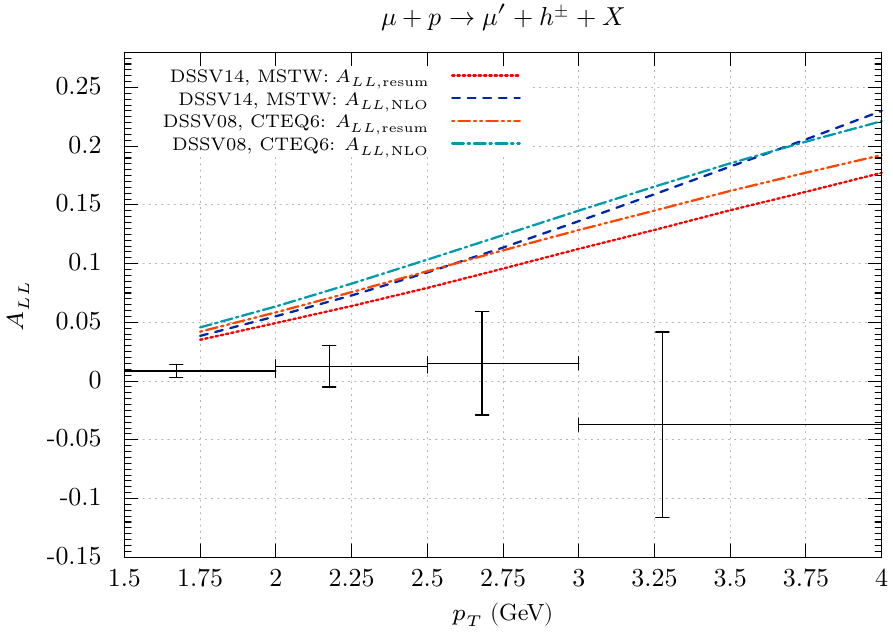}}   
  \end{center}
  \caption{\label{asymmetry-COMPASS} {\it Double-longitudinal spin asymmetries $A_{LL}$ for (a) a deuteron and (b) a proton target for 
  COMPASS kinematics with the full rapidity range $-0.1\leq \eta \leq 2.38$. In both cases, we show the NLO and matched resummed results for two different 
  sets of parton distributions (see text). The asymmetries include the resolved contributions at NLO; for illustration we also show in (a) the resummed 
  asymmetry without the resolved contributions. The theoretical results are compared to the recent COMPASS data~\cite{Adolph:2015hta}.}}
\end{figure}
We now investigate the double-longitudinal spin asymmetry for single-inclusive hadron production 
with a deuteron or a proton target. It is given by the ratio of the spin-dependent and the spin-averaged 
cross sections defined in Eq.~(\ref{lNhel}):
\begin{align}
 A_{LL}=\frac{d\Delta \sigma}{d \sigma} \,.
\end{align}
We include the NLO resolved photon contributions, so that at the present stage the ``resummed'' spin 
asymmetry is given by
\begin{align}
 A_{LL,\text{resum}}=\frac{ d\Delta \sigma_{\text{dir,resum}} + d\Delta\sigma_{\text{res,NLO}}}{d\sigma_{\text{dir,resum}} + d\sigma_{\text{res,NLO}}},
\end{align}
while the NLO one is as usual 
\begin{align}
 A_{LL,\text{NLO}}=\frac{ d\Delta \sigma_{\text{dir,NLO}} + d\Delta\sigma_{\text{res,NLO}}}{d\sigma_{\text{dir,NLO}} + d\sigma_{\text{res,NLO}}}.
\end{align}
Our results are shown in Figs.~\ref{asymmetry-COMPASS} (a) and (b).
The different size of the resummation effects for the polarized and unpolarized cross sections that we found
in Fig.~\ref{compare_unpol-pol} clearly implies that the resummed threshold logarithm contributions do not 
cancel in the double-spin asymmetry. Indeed, as Fig.~\ref{asymmetry-COMPASS} shows, for our default
sets of parton distributions the deuteron asymmetry is reduced by almost a factor of two at high $p_T$, 
when going from NLO to the resummed case. For a proton target, there also is a substantial, albeit
somewhat less dramatic, decrease. We also plot in the figure the corresponding results obtained by
using the unpolarized and polarized parton distributions of Refs.~\cite{CTEQ6M5} (CTEQ6.5M) 
and~\cite{DSSV} (DSSV2008), respectively. For these, the main trends are
qualitatively similar, although the redcuction of $A_{LL}$ is slightly less pronounced. This is
likely due to the fact that the DSSV2008 set has a smaller gluon helicity distribution $\Delta g$,
so that the Compton process $\gamma q\to qg$ dominates, which has a positive partonic
spin asymmetry and receives similar resummation effects in the unpolarized and the
polarized case. We finally note that for the case of a deuteron target in Fig.~\ref{compare_unpol-pol} we
also show the asymmetry based on the direct contributions alone. Evidently, this asymmetry is
much smaller, expressing the fact that resolved contributions are likely very important for the polarized
cross section. 

As stated in the Introduction, COMPASS has recently presented data for the spin asymmetries
for deuteron and proton targets~\cite{Adolph:2015hta}. The data (combined for the full rapidity range
$-0.1\leq \eta \leq 2.38$ and summed over hadron charges) are shown in Fig.~\ref{asymmetry-COMPASS} 
in comparison to our theoretical results. As one can see, while the asymmetries for deuterons are in marginal
agreement, the very small asymmetry seen by COMPASS for protons is incompa\-tible with any of our
predictions. It is worth mentioning that, as shown in Ref.~\cite{Adolph:2015hta}, this problem appears to 
be especially pronounced in the rapidity range $-0.1\leq \eta \leq 0.45$ and for positively charged hadrons. 
While the higher order resummed corrections that we have included ameliorate the situation, 
they are clearly not sufficient. Given the rather large decrease of the spin asymmetry generated by 
resummation of the direct contributions, it is arguably not possible to draw any reliable conclusions 
form this observation before also the resummation for the resolved part of the cross sections has 
been carried out. It appears unlikely, however, that the resolved contribution and its resummation 
will bring the data and theoretical results into good agreement since they affect the asymmetries
for both targets in similar ways. If, for instance, the polarized resolved contribution were so large and negative 
that the proton data could be accommodated, the description of the deuteron asymmetry would
vastly deteriorate~\cite{Julius}. 

\section{Conclusions and outlook \label{sec5}}
We have studied the impact of threshold resummation at next-to-leading logarithmic level 
on the spin-dependent cross section for $\gamma N \rightarrow h X$ at high transverse 
momentum $p_T$ of the hadron $h$, and on the resulting double-longitudinal spin asymmetry $A_{LL}$. 
For the present work, we have implemented the resummation only for the direct contribution
to the cross section. For the kinematics relevant for the COMPASS experiment we find
that the spin-dependent cross section receives much smaller enhancements by resummation
than the spin-averaged one treated in Ref.~\cite{MelaniesPaper}. As a result, threshold effects do 
not cancel in the double-spin asymmetry, and the prediction for $A_{LL}$ decreases when resummation 
is taken into account. Definite conclusions about the impact of resummation on the spin asymmetry
will become possible only when also the resummation for the resolved component has been 
carried out, which we plan to do in future work. Only then will an extraction of the proton's
gluon helicity distribution $\Delta g$ become meaningful. We also note that the scale dependence of 
the perturbative cross section remains uncomfortably large even when resummation for the direct
piece is taken into account. In order to improve this it may, eventually, be necessary to extend
resummation to next-to-next-to-leading logarithmic level, following the techniques developed
in Ref.~\cite{Hinderer:2014qta}.

Comparison to the recent COMPASS data~\cite{Adolph:2015hta} shows that the theoretically
predicted spin asymmetries fail to reproduce the data well. Especially for the proton target the
data show a nearly vanishing asymmetry, while the theoretical result appears to be always 
clearly positive. In fact, it is worth stressing that each of the theoretical results shown in 
Fig.~\ref{asymmetry-COMPASS} predicts a {\it larger} spin asymmetry for the proton than for 
the deuteron, in contrast to the trend seen in the data. This feature of the theoretical predictions
is likely no accident, as a simple study of the LO direct contributions shows~\cite{Julius}.
Clearly, future work is needed in order to clarify in how far the leading-twist perturbative-QCD framework
can accommodate a larger spin asymmetry for $\mu d \rightarrow \mu' h X$ than for
$\mu p \rightarrow \mu' h X$.

\begin{acknowledgments}
We are grateful to Y. Bedfer, F. Kunne, M. Levillain, C. Mar\-chand, M. Pfeuffer, 
and J. Steiglechner for useful discussions. This work was supported by the 
``Bundesministerium f\"{u}r Bildung und Forschung'' (BMBF) grants no. 05P12WRFTE and 05P12VTCTG. 
\end{acknowledgments}
\begin{appendix}

\section{Coefficients $\Delta C_{\gamma b \rightarrow cX}$\label{AppA}}

\renewcommand{\theequation}{A.\arabic{equation}}
\setcounter{equation}{0} 

In order to present our results for the $\Delta C_{\gamma b \rightarrow cX}$ in compact form,
we define
\begin{align}
 \rho^{(A)}_{q \gamma} &= 4 \gamma_E + 4 \ln 2 , \nonumber \\[2mm]
 \rho^{(F)}_{q \gamma} &= -3+4\gamma_E + 4 \ln\left( 2(1-v)\right), \nonumber \\[2mm]
 \rho^{(A)}_{g \gamma} &= 4 \gamma_E + 4 \ln\left( 2(1-v)\right)  , \nonumber \\[2mm]
 \rho^{(F)}_{g \gamma} &= -3 + 4\gamma_E + 4 \ln 2 ,
\end{align}
where $\gamma_E$ is the Euler constant. For the Compton process $\gamma q\rightarrow q g$ we then have
\begin{align}
 & \! \Delta C_{\gamma q\rightarrow q g} \! = b_0 \pi \ln \frac{\mu_r^2}{\hat{s}}   + \frac{C_F}{4} 
 \ln \frac{\mu_{ff}^2}{\hat{s}}  \left( \rho^{(A)}_{q \gamma} -3 \right)
  \nonumber \\[2mm]
  & \hspace{0.2cm} + \frac{C_F}{4}  \ln  \frac{\mu^2_{fi}}{\hat{s}} \rho^{(F)}_{q \gamma} + 
  \frac{1}{18} \left( 2C_A - 5 N_f \right) + \frac{1}{4} b_0 \pi \rho^{(A)}_{q \gamma} \nonumber \\[2mm]
   & \hspace{0.2cm}  + \frac{C_A^2 -2}{32 C_A} \left( \rho^{(A)}_{q \gamma} \right)^2 + \frac{\ln v}{4 C_A} 
   \left( \rho^{(A)}_{q \gamma} - \ln v  \right) \nonumber \\[2mm]
   & \hspace{0.2cm} +\frac{\pi^2}{4 C_A} \frac{2v-1}{v(v-2)} + \frac{7}{4 C_A} + \frac{\pi^2 C_F}{3} \nonumber \\
  & \hspace{0.2cm} + \frac{ \ln\left( 1-v \right)}{2 C_A v(v-2)}  \left\{ \ln\left( \frac{\sqrt{1-v}}{v} \right)  \left(4v-v^2-1 \right)  
  \right. \nonumber \\[2mm]
  & \hphantom{  \hspace{0.2cm} + \frac{ \ln\left( 1-v \right)}{2 C_A v(v-2)} \left\{ \right. } 
  \left. + \frac{1}{2} \left[ 1 \! -3 C_A^2 +2v + \rho^{(A)}_{q \gamma} \left( 2v-v^2  \right) \!  \right] \right\}, \nonumber \\
\end{align}
where $C_F=4/3, C_A=3$. For the process $\gamma q \rightarrow g q$ with an observed gluon, 
\begin{align}
 & \! \Delta C_{\gamma q\rightarrow g q} \! =  b_0 \pi \ln  \frac{\mu_r^2}{\hat{s}}  +  \ln \frac{\mu_{ff}^2}{\hat{s}}   
 \left( \frac{C_A  }{4} \rho^{(A)}_{q \gamma}  -  b_0 \pi   \right)  \nonumber \\[2mm]
    & \hspace{0.2cm} + \! \frac{C_F}{4} \rho^{(F)}_{q \gamma} \! \ln \! \frac{\mu^2_{fi}}{\hat{s}}  
    \! + \bigg( \! \frac{C_F}{2} + C_A \! \bigg) \! {\left(\! \frac{\rho^{(A)}_{q \gamma}}{4} \! \right)\!} ^2 \! + 
    \! \frac{\pi^2 \left( v^2-6v+2 \right)}{12 C_A \left( v^2-1 \right)} \nonumber \\[2mm]
    & \hspace{0.2cm} + \frac{1}{12} \Big\{ \frac{ 3 C_F}{4} (3 \rho^{(A)}_{q \gamma}-28) + \pi^2  (4 C_A +C_F)\Big\}
      \nonumber \\[2mm]
   & \hspace{0.2cm} + \frac{\ln v }{4} \Big\{  C_F (2\ln v+3) - C_A \big( 2 \ln(1-v) +  \rho^{(A)}_{q \gamma} \big) \nonumber \\[2mm]
   & \hphantom{\hspace{0.2cm} + \frac{\ln v }{4} \bigg\{} 
    - \frac{1}{C_A \left(v^2-1\right)} \Big[- v(v-2) \ln v + 3 C_A C_F (v^2+1)  \nonumber \\[2mm]
   & \hphantom{\hspace{0.2cm} + \frac{\ln v }{4} \Big\{ - \frac{1}{C_A \left(v^2-1\right) } } 
     + 2 \ln (1-v)(1-2v)+2v \Big] \Big\} \nonumber \\[2mm]
   & \hspace{0.2cm} + \frac{C_A}{4 }  \ln \left( 1-v \right) \left[ \rho^{(A)}_{q \gamma} +  \ln \left(1-v \right) \right].  
\end{align}
Finally, for photon-gluon fusion $\gamma g \rightarrow q \bar{q}$, we find
\begin{align}
  &  \! \Delta C_{\gamma g \rightarrow q \bar{q}} \! = b_0 \pi \ln  \frac{\mu_r^2}{\mu_{fi}^2} +  
  \ln \frac{\mu_{fi}^2}{\hat{s}}   \frac{C_A}{4} \rho^{(A)}_{g \gamma}  
   +   \frac{C_F}{4}  \ln  \frac{\mu_{ff}^2}{\hat{s}}  \rho^{(F)}_{g \gamma} \nonumber \\[2mm]
   & \hspace{0.2cm} + \frac{1}{6} \bigg\{ C_A \left[ \frac{3}{8} \Big( \rho^{(F)}_{g \gamma} + 3 \Big)^2 + \! \!\pi^2 \right]
   \nonumber \\[2mm]
   & \hphantom{\hspace{0.2cm} + \frac{1}{6} \bigg\{ }  
   + C_F \left[ \frac{9}{8}\Big( \rho^{(F)}_{g \gamma} - \frac{19}{3} \Big) + \!\frac{5}{2} \pi^2 \right]
   \bigg\} \nonumber \\[2mm]
   & \hspace{0.2cm} + \frac{\ln v}{8 C_A}  \bigg\{ \frac{3 C_A^2 \left( 1-2v \right) +2v\left(1+2v\right)-3 }{v^2+(1-v)^2}
       \nonumber \\[2mm]
   & \hphantom{ \hspace{0.2cm} + \frac{\ln v}{8 C_A}  \bigg\{ } -2 C_A^2 \rho^{(A)}_{g \gamma} + 
   6 C_A C_F \bigg\} \nonumber \\[2mm]
   & \hspace{0.2cm} + \ln \left(1-v \right) \left\{ \frac{3 C_A^2 v^2 -v(v+2)}{v^2+(1-v)^2} +
   C_A^2 \Big(  \rho^{(F)}_{g \gamma} +3  \Big) \right\} \nonumber \\[2mm]
   & \hspace{0.2cm} -\frac{ \ln^2 v}{4 C_A} \left\{ \frac{1+v^2}{v^2+(1-v)^2} - C_A^2 \right\} \nonumber \\[2mm]
   & \hspace{0.2cm} -\frac{\ln^2 \left(1-v \right)}{4C_A}  \left\{ \frac{1+(1-v)^2}{v^2+(1-v)^2} - C_A^2 \right\}.
\end{align}
We note that 
$ \Delta C_{\gamma g \rightarrow q \bar{q}}$ is identical to the corresponding coefficient $C_{\gamma g \rightarrow q \bar{q}}$
in the unpolarized case, which was given in~\cite{MelaniesPaper}. 

\section{Radiative exponents and their expansion to NLL\label{AppB}}

\renewcommand{\theequation}{B.\arabic{equation}}
\setcounter{equation}{0}

The perturbative expansion of the function $A_i$ to the required order is given by 
\begin{align}
 A_i \left( \alpha_s \right) = & \frac{\alpha_s}{\pi} A_{i}^{(1)} + \left( \frac{\alpha_s}{\pi} \right)^2 A_{i}^{(2)} + 
 \mathcal{O}(\alpha_s ^3)  \nonumber \\[2mm]
	= & C_i \left( \frac{\alpha_s}{\pi} + \frac{1}{2} K \left( \frac{\alpha_s}{\pi} \right)^2 \right)+ 
 \mathcal{O}(\alpha_s ^3) , 
	\end{align}
where $C_f=C_F=4/3$ for a quark and $C_f=C_A=3$ for a gluon, and where
\begin{align}
 K=C_A \left( \frac{67}{18} - \frac{\pi^2}{6} \right) - \frac{5}{9} N_f,
\end{align}
with $N_f$ the number of flavors. The quark and gluon field anomalous dimensions $\gamma_i$  
and the leading terms $\gamma_{ii}$ of the diagonal splitting functions read to one-loop order 
\cite{LaenenJuni1998}:
\begin{align}
 \gamma_{q}(\alpha_s) & = \frac{3}{4} C_F \frac{\alpha_s}{\pi} \text{,} 
 &\gamma_{qq}(N, \alpha_s) = - \left( \ln N - \frac{3}{4} \right) C_F \frac{\alpha_s}{\pi} \text{,} \nonumber \\[2mm]
 \gamma_{g}(\alpha_s) & = b_0 \alpha_s \text{,}
 & \gamma_{gg} (N, \alpha_s) = - \left( C_A \ln N - \pi b_0 \right)\frac{\alpha_s}{\pi} \text{,}
\end{align}
where $b_0=(11 C_A - 4 T_R N_f)/(12 \pi)$, with $T_R=1/2$.

We next present the next-to-leading logarithmic expansions of $\ln  \Delta_i^{N}$ and
$\ln J_d^{N}$ in Eqs. (\ref{ln_Delta}) and (\ref{ln_J}), respectively. Defining 
\begin{align}
 \lambda\equiv b_0 \alpha_s \ln (N{\mathrm{e}}^{\gamma_E}),
\end{align}
we have~\cite{Catani1998,Sterman2000,MelaniesPaper}: 
\begin{align}
   \ln  \Delta_i^N  \left(\hat{s},\mu_{fi},\mu_r \right)=&  \ln N \,h_i^{(1)}(\lambda) 
    +h_i^{(2)}\left( \lambda,\frac{\hat{s}}{\mu_r^2} , \frac{\hat{s}}{\mu_{fi}^2}\right) ,     
\label{NLL_exponents1}
\end{align}
and
 \begin{align}
    \ln J_i^{N}(\hat{s},\mu_r) =& \ln N \, 
    f_i^{(1)} (\lambda) + f_i^{(2)} \left(\lambda, \frac{\hat{s}}{\mu_r^2} \right) .
    \label{NLL_exponents2}
  \end{align}
These exponents are universal in the sense that they depend only on the parton considered, but not on the 
overall subprocess. The functions $h_{i}^{(1)}$  and $f_{i}^{(1)}$ collect all leading logarithmic terms 
$\alpha_s^k \ln^{k+1} N$ in the exponent, while the $h_{i}^{(2)}$  and $f_{i}^{(2)}$ produce next-to-leading 
logarithms $\alpha_s^k \ln^k N$. They read~\cite{Catani1998}
\begin{align}
 h_{i}^{(1)} & \left( \lambda \right) = 
 \frac{A_{i}^{(1)}}{2 \pi b_0 \lambda} \left[ 2 \lambda +  \left( 1- 2 \lambda \right) \ln{ 
 \left( 1- 2 \lambda \right)}\right] \displaybreak[1] \text{,} \\[2mm]
 h_{i}^{(2)} & \left( \lambda , \frac{Q^2}{\mu_{r}^{2}} , \frac{Q^2}{\mu_{f}^{2}} \right) = 
 - \frac{A_{i}^{(2)}}{2 \pi^2 b_{0}^2} \left[ 2 \lambda + \ln{ \left( 1- 2 \lambda \right)}\right] \nonumber \\[2mm]
  & + \frac{A_{i}^{(1)} b_1 } {2 \pi b_0^3} \left[ 2 \lambda +  \ln{ \left( 1- 2 \lambda \right)}+ 
  \frac{1}{2} \ln^2{ \left( 1- 2 \lambda \right)} \right] \nonumber  \\[2mm]
 &  - \frac{A_{i}^{(1)} } { \pi b_{0}} \lambda \ln{ \frac{Q^2}{\mu_f^2} } + 
 \frac{A_{i}^{(1)}} {2 \pi b_0} \left[ 2 \lambda +  \ln{ \left( 1- 2 \lambda \right)} \right] \ln {\frac{Q^2}{\mu_r^2}}
 \text{,} 
\end{align}
with $b_1 = (17 C_A^2 - 5 C_A N_f - 3 C_F N_f ) / (24 \pi^2) $. Furthermore,
\begin{align}
 f_{i}^{(1)} \left( \lambda \right) =  & h_{i}^{(1)}  \left( \lambda/2 \right)- h_{i}^{(1)}  \left( \lambda \right) \nonumber \\[2mm]
 f_{i}^{(2)}  \left( \lambda , \frac{Q^2}{\mu_{r}^{2}}  \right) = &
 2h_{i}^{(2)} \left(\frac{\lambda}{2} , \frac{Q^2}{\mu_{r}^{2}} , 1 \right) -h_{i}^{(2)} \left( \lambda , \frac{Q^2}{\mu_{r}^{2}} , 1 \right)
   \nonumber \\[2mm]
&  + \frac{B_{i}^{(1)}} {2 \pi b_0} \ln{ \left( 1-\lambda \right)}  ,
\end{align}
where $B_i^{(1)}=-2 \gamma_i^{(1)}$.

\end{appendix}

\newpage
\end{document}